\documentclass{article}%
\usepackage{amsmath}
\usepackage{graphicx}%
\usepackage{amsfonts}%
\usepackage{amssymb}
\newtheorem{theorem}{Theorem}

\newtheorem{definition}[theorem]{Definition}

\newtheorem{proposition}[theorem]{Proposition}

\newenvironment{proof}[1][Proof]{\noindent\textbf{#1.} }{\ \rule{0.5em}{0.5em}}
\begin{document}

\title{Is Quantum Mechanics An Island In Theoryspace?}
\author{Scott Aaronson\thanks{University of California, Berkeley. Email:
aaronson@cs.berkeley.edu. Supported by an NSF Graduate Fellowship, by NSF ITR
Grant CCR-0121555, and by the Defense Advanced Research Projects Agency (DARPA).}}
\date{}
\maketitle

\begin{abstract}
This recreational paper investigates what happens if we change quantum
mechanics in several ways. \ The main results are as follows. \ First, if we
replace the $2$-norm by some other $p$-norm, then there are no nontrivial
norm-preserving linear maps. \ Second, if we relax the demand that norm be
preserved, we end up with a theory that allows rapid solution of $\mathsf{PP}%
$-complete problems (as well as superluminal signalling). \ And third, if we
restrict amplitudes to be real, we run into a difficulty much simpler than the
usual one based on parameter-counting of mixed states.

\end{abstract}

\section{Introduction\label{INTRO}}

\begin{quote}
``It is striking that it has so far not been possible to find a logically
consistent theory that is close to quantum mechanics, other than quantum
mechanics itself.''\ ---Steven Weinberg, \textit{Dreams of a Final Theory} \cite{weinberg2}
\end{quote}

The title of this paper should be self-explanatory, but if it isn't:
\textquotedblleft theoryspace\textquotedblright\ is the space of logically
conceivable physical theories, with two theories close to each other if they
differ in few respects. \ An \textquotedblleft island\textquotedblright\ in
theoryspace is a natural and interesting theory, whose neighbors are all
somehow perverse or degenerate.\footnote{A bit of pedantry: a physicist might
call the neighbors of quantum mechanics I'll discuss \textquotedblleft
inconsistent,\textquotedblright\ since they contradict auxiliary assumptions
that the physicist considers obvious. \ I'll stick to milder epithets like
\textquotedblleft perverse.\textquotedblright} \ The Standard Model isn't an
island, because we don't know any compelling (non-anthropic) reason why the
masses and coupling constants should have the values they do.\footnote{More
pedantry: whether a theory is an island is therefore a function of our
knowledge, not just of the theory itself.} \ Likewise, general relativity is
probably not an island, because of alternatives such as the Brans-Dicke theory.

To many physicists, however, quantum mechanics \textit{does} seem like an
island: change any one aspect, and the whole structure collapses. \ This view
is buttressed by three types of results:

\begin{enumerate}
\item[(1)] \textbf{\textquotedblleft Derivations\textquotedblright\ of the
}$\left\vert \mathbf{\psi}\right\vert ^{2}$\textbf{ probability rule.}
\ Gleason's Theorem \cite{gleason}\ shows that, in a Hilbert space of dimension
$3$ or higher, the usual quantum probability rule is the only one consistent
with a requirement of noncontextuality. \ Deutsch \cite{deutsch} and
Zurek \cite{zurek} derived the rule from other assumptions.

\item[(2)] \textbf{Arguments for complex amplitudes.} \ If $f\left(  n\right)
$\ is the number of real parameters needed to specify an $n$-dimensional mixed
state, then only when amplitudes are complex numbers does $f\left(  n_{A}%
n_{B}\right)  =f\left(  n_{A}\right)  f\left(  n_{B}\right)  $\ (since
$f\left(  n\right)  =n^{2}$). \ With real amplitudes, $f\left(  n\right)
=n\left(  n+1\right)  /2$\ and thus $f\left(  n_{A}n_{B}\right)  >f\left(
n_{A}\right)  f\left(  n_{B}\right)  $. \ With quaternionic amplitudes,
$f\left(  n\right)  =2n^{2}-n$\ and thus $f\left(  n_{A}n_{B}\right)
<f\left(  n_{A}\right)  f\left(  n_{B}\right)  $. \ Caves, Fuchs, and
Schack \cite{cfs} exploited this observation to show that a \textquotedblleft
quantum de Finetti Theorem\textquotedblright\ (which justifies Bayesian
reasoning) works only if amplitudes are complex.\ \ Hardy \cite{hardy} also
made essential use of the observation in his derivation of quantum mechanics
from \textquotedblleft five simple axioms.\textquotedblright

\item[(3)] \textquotedblleft\textbf{Perverse\textquotedblright\ consequences
of nonlinearity.} \ After Weinberg \cite{weinberg}\ proposed nonlinear variants
of the Schr\"{o}dinger equation, Gisin \cite{gisin}\ and
Polchinski \cite{polchinski}\ independently observed that almost all such
variants would allow superluminal signalling. \ Later Abrams and
Lloyd \cite{al} argued that a \textquotedblleft nonlinear quantum
computer\textquotedblright\ could solve $\mathsf{NP}$-complete and even
$\mathsf{\#P}$-complete problems in polynomial time.\footnote{Abrams and Lloyd
claimed furthermore that their nonlinear algorithms are robust against small
errors. \ This claim does not withstand detailed scrutiny; whether nonlinear
quantum computers can solve $\mathsf{NP}$- and $\mathsf{\#P}$-complete
problems robustly therefore remains an intriguing open problem. \ On the other
hand, if arbitrary $1$-qubit nonlinear gates can be implemented without error,
then even $\mathsf{PSPACE}$-complete problems can be solved in polynomial
time. \ This result is tight, since nonlinear quantum computers can also be
simulated in $\mathsf{PSPACE}$. \ These claims will be proved in another paper.}
\end{enumerate}

This paper won't attempt another axiomatic derivation like that of
Hardy \cite{hardy}---its more modest goal is just to stroll through quantum
mechanics' neighborhood of theoryspace. \ All mathematical results in this
paper are trivial. \ So why write it then? \ Apart from the fact that
triviality never stopped a quantum philosopher before, I hope to make a point:
that if you change quantum mechanics in the most obvious ways, you'll run into
problems that have nothing to do with the subtleties of\ contextuality,
locality, or entanglement. \ Even in \textquotedblleft quantum mechanics
lite\textquotedblright---where there are no mixed states, no tensor products,
and no intermediate measurements, just vectors representing probabilities that
get mapped to other vectors---you'll need to worry about conservation of
probability, and about closure properties of the allowed vector maps.

I \textit{won't} make this point regarding nonlinear quantum mechanics, for
the simple reason that there it seems false. \ Contrary to what I originally
thought, one can define a large, natural class of discrete norm-preserving
nonlinear gates. \ This class includes \textquotedblleft Weinberg
gates\textquotedblright\ such as%
\[
W\left(
\begin{array}
[c]{c}%
x\\
y
\end{array}
\right)  =\left(
\begin{array}
[c]{c}%
x\\
e^{iy}y
\end{array}
\right)  ,
\]
as well as \textquotedblleft polynomial gates\textquotedblright\ such as%
\[
G\left(
\begin{array}
[c]{c}%
x\\
y
\end{array}
\right)  =\left(
\begin{array}
[c]{c}%
x^{2}-\left(  y^{\ast}\right)  ^{2}\\
2\operatorname{Re}xy
\end{array}
\right)  .
\]
Since $\left\Vert G\left(  v\right)  \right\Vert _{2}=\left\Vert v\right\Vert
_{2}^{2}$, the gate $G$ preserves the $2$-norm of $v$ provided $\left\Vert
v\right\Vert _{2}=1$. \ As far as I can tell, any argument for the
implausibility of $W$ or $G$ needs to be based on \textit{physical} effects,
such as superluminal signalling or efficient solubility of $\mathsf{NP}%
$-complete problems.

The paper is (not very well) organized as follows. \ Section \ref{PNORM}%
\ shows that when $p\neq2$, the only $p$-norm-preserving linear
transformations are permutations of diagonal matrices. \ In other words, if
you want to base quantum mechanics on a $p$-norm other than the $2$-norm, then
you'll need to include some sort of \textquotedblleft manual
normalization.\textquotedblright\ \ However, manual normalization brings with
it most of the hazards of nonlinearity: superluminal signalling,
distinguishability of non-orthogonal states, and polynomial-time solubility of
\textquotedblleft obviously hard\textquotedblright
\ problems.\footnote{$\mathsf{NP}$-complete problems are obviously hard;
factoring and graph isomorphism are not.} \ Section \ref{POSTBQP} addresses
the last point in detail, by using the concept of \textit{postselection} to
study the computational power of alternative quantum theories. \ The
punchline, which might be of independent interest to computer scientists, is
that all the alternative theories considered have at least the power of the
complexity class $\mathsf{PP}$,\footnote{See www.cs.berkeley.edu/\symbol{126}%
aaronson/zoo.html for definitions of over $370$ complexity classes.} and many
have \textit{exactly} the power of $\mathsf{PP}$.

Finally, Section \ref{REAL}\ gives an argument for why amplitudes are complex
rather than real, that has nothing to do with the parameter-counting arguments
of Refs. \cite{cfr,cfs,hardy}. \ Unfortunately, my argument says nothing about
why amplitudes are complex rather than quaternionic.

\section{Other $p$-Norms\label{PNORM}}

\begin{quote}
\textquotedblleft Addition in proof: More careful considerations show that the
probability is proportional to the square of the [amplitude] $\Phi_{nrm}%
$.\textquotedblright\ ---Max Born \cite{born}, in a footnote to his 1926 paper
introducing the probability interpretation (the main text says the probability
is proportional to $\Phi_{nrm}$ itself)
\end{quote}

No doubt about it: the $2$-norm is special. \ The Pythagorean Theorem,
Fermat's Last Theorem, and least-squares regression all involve properties of
a sum of squares that fail for a sum of cubes or of any other power.\ \ Still,
given that classical probability theory is based on the $1$-norm and quantum
mechanics on the $2$-norm, it's natural to wonder what singles out $1$ and
$2$. \ What happens if we try to base a theory on the $p$-norm\footnote{The
main reason for restricting attention to $p$-norms is their behavior under
tensor products: disregarding zany functions that depend on the Axiom of
Choice, if $f\left(  \alpha\beta\right)  =f\left(  \alpha\right)  f\left(
\beta\right)  $\ for all $\alpha,\beta$\ then $f\left(  \alpha\right)  $ must
have the form $\left\vert \alpha\right\vert ^{p}$. \ However, it might be
interesting to consider theories where the probability of measuring a basis
state $\left\vert x\right\rangle $\ depends on \textit{all} amplitudes, not
just that of $\left\vert x\right\rangle $.} for some other $p$? \ In this
section I'll explain why the $2$-norm is the only $p$-norm that permits
nontrivial norm-preserving linear maps.\footnote{When $p=0$\ all linear maps
are norm-preserving, but they have no effect because all outcomes of a
measurement are always equiprobable. \ When $p=\infty$\ only generalized
diagonal matrices are norm-preserving, as in the case $2<p<\infty$. \ I refuse
even to discuss the case $p<0$.}

It's easiest to start with real amplitudes and then generalize to complex
ones.\ \ We want to know which matrices $A\in\mathbb{R}^{n\times n}$\ have the
property that for all vectors $x$, $\left\Vert Ax\right\Vert _{p}=\left\Vert
x\right\Vert _{p}$, where $\left\Vert \cdot\right\Vert _{p}$\ denotes the
$p$-norm. \ We can gain some intuition by counting constraints. \ When
$p=1$\ \textit{and} we restrict our attention to $x$ with nonnegative entries,
we obtain the set of stochastic matrices, or nonnegative matrices that satisfy
$n$ linear constraints. \ When $p=2$, we obtain the set of orthogonal
matrices, or those $A=\left(  a_{jk}\right)  $\ such that%
\begin{equation}
\sum_{j=1}^{n}a_{jk}a_{kl}=\delta_{kl} \label{orthog}%
\end{equation}
for all $k,l$. \ Equation \ref{orthog} imposes $n\left(  n+1\right)
/2$\ quadratic constraints on $A$, cutting the number of parameters needed to
specify $A$ roughly in half. \ Continuing, when $p=3$\ we expect order $n^{3}%
$\ cubic constraints, when $p=4$, order $n^{4}$\ quartic constraints, and so
on. \ That the number of constraints exceeds the number of parameters for
$p>2$\ makes us suspect that $p=2$\ is the \textquotedblleft end of the line.\textquotedblright

But that's not a rigorous argument, because we know there are matrices that
are norm-preserving for all $p$: the generalized diagonal matrices (that is,
products of permutation matrices and diagonal matrices). \ To show that these
are the only norm-preserving matrices, first let $p$ be an even integer
greater than $2$. \ Then letting $x=\left(  x_{j}\right)  $, the requirement%
\begin{equation}
\sum_{j=1}^{n}x_{j}^{p}=\sum_{j=1}^{n}\left(  \sum_{k=1}^{n}a_{jk}%
x_{k}\right)  ^{p} \label{porthog}%
\end{equation}
for all $x$ implies that the left- and right-hand sides are identical as
formal polynomials, and therefore (among other constraints) that%
\[
\sum_{j=1}^{n}a_{jk}^{p-2}a_{jl}^{2}=\delta_{kl}%
\]
for all $k,l$. \ This in turn implies that for all $j$\ and $k\neq l$, either
$a_{jk}=0$\ or $a_{jl}=0$. \ But since every column must contain nonzero
entries by the constraint $\sum_{j}a_{jk}^{p}=1$, this implies that $A$ is a
generalized diagonal matrix.

Next let $p$ be an odd positive integer. \ We claim that, so long as
$x_{1},\ldots,x_{n}$ are nonnegative, the entries of $Ax$ never change sign.
\ Clearly there exist $s_{1},\ldots,s_{n}\in\left\{  -1,1\right\}  $\ such
that%
\[
\sum_{j=1}^{n}x_{j}^{p}=\sum_{j=1}^{n}s_{j}y_{j}^{p}%
\]
as formal polynomials, where $y_{j}=\sum_{k=1}^{n}a_{jk}x_{k}$. \ Suppose by
contradiction that, keeping all $x_{j}$'s nonnegative, we could make
$\operatorname*{sgn}\left(  y_{j}\right)  s_{j}=-1$ for some $j$, where
$\operatorname*{sgn}\left(  y_{j}\right)  $ is $0$ if $y_{j}=0$ and$\ y_{j}%
/\left\vert y_{j}\right\vert $ otherwise. \ Then%
\[
\sum_{j=1}^{n}\operatorname*{sgn}\left(  y_{j}\right)  y_{j}^{p}=\sum
_{j=1}^{n}x_{j}^{p}=\sum_{j=1}^{n}s_{j}y_{j}^{p}%
\]
as formal polynomials, which implies that%
\[
\sum_{\operatorname*{sgn}\left(  y_{j}\right)  s_{j}=-1}\operatorname*{sgn}%
\left(  y_{j}\right)  y_{j}^{p}=0.
\]
Since every term in the above sum is nonnegative, we have $y_{j}^{p}=0$\ for
all $j$\ such that $\operatorname*{sgn}\left(  y_{j}\right)  s_{j}=-1$, which
implies that $a_{jk}=0$\ for all $j,k$\ such that $\operatorname*{sgn}\left(
y_{j}\right)  s_{j}=-1$, contradiction.

Since the entries of $Ax$\ never change sign when $x$ is nonnegative, it
follows that in each row of $A$, all entries have the same sign. \ So if we
define a new matrix $B$ by $b_{jk}=\left\vert a_{jk}\right\vert $, then $B$
also has the property that $\left\Vert Bx\right\Vert _{p}=\left\Vert
x\right\Vert _{p}$\ for all $x$. \ But then when $p\geq3$, the same reasoning
from the case of even $p$ implies that $B$ is generalized diagonal, which
implies that $A$ was generalized diagonal as well. \ When $p=1$, $B$ is
stochastic, and it is easily checked that the only stochastic matrices that
preserve the $1$-norm of all vectors (not just nonnegative ones) are
permutation matrices.

Finally, let $p>0$ be an arbitrary real that is not an integer. \ Let
$\widetilde{x}_{j}=\left\vert x_{j}\right\vert ^{p}$; then%
\[
\sum_{j=1}^{n}\widetilde{x}_{j}=\sum_{j=1}^{n}\left\vert \sum_{k=1}^{n}%
a_{jk}\widetilde{x}_{k}^{1/p}\right\vert ^{p}%
\]
for all $\widetilde{x}_{1},\ldots,\widetilde{x}_{n}$. \ It follows that there
exist $s_{1},\ldots,s_{n}\in\left\{  -1,1\right\}  $\ such that%
\[
\sum_{j=1}^{n}\widetilde{x}_{j}=\sum_{j=1}^{n}\left(  s_{j}\sum_{k=1}%
^{n}a_{jk}\widetilde{x}_{k}^{1/p}\right)  ^{p}%
\]
as formal functions. \ But this implies that $A$ is generalized diagonal,
since otherwise the right-hand side could never be simplified to a linear
function in the $\widetilde{x}_{j}$'s.

So much for real amplitudes. \ When we generalize to complex amplitudes
$x_{j}\in\mathbb{C}$, there are two defensible choices: letting $x_{j}%
=\alpha_{j}+i\beta_{j}$, we could require either $\sum_{j=1}^{n}\left(
\left\vert \alpha_{j}\right\vert ^{p}+\left\vert \beta_{j}\right\vert
^{p}\right)  =1$ or $\sum_{j=1}^{n}\left\vert x_{j}\right\vert ^{p}=1$, where
$\left\vert x_{j}\right\vert =\sqrt{\alpha_{j}^{2}+\beta_{j}^{2}}$ as usual.
\ Under the first choice, we can consider $\alpha_{1},\ldots,\alpha_{n}%
,\beta_{1},\ldots,\beta_{n}$ as a vector of $2n$\ reals and $A$ as a
$2n\times2n$\ matrix; then the results from the real-amplitude case
immediately imply that $A$ is generalized diagonal. \ Under the second choice,
we can choose an $x_{l}\neq0$\ and replace it by $e^{i\theta}x_{l}$, holding
all other $x_{k}$'s fixed. \ Then since $\sum_{j=1}^{n}\left\vert
x_{j}\right\vert ^{p}$\ remains constant as we vary $\theta$,%
\[
\sum_{j=1}^{n}\left\vert y_{j}\right\vert ^{p}=\sum_{j=1}^{n}\left\vert
a_{jl}e^{i\theta}x_{l}+\sum_{k\neq l}a_{jk}x_{k}\right\vert ^{p}%
\]
must also remain constant. \ But when $p\neq2$, this is possible only if for
all $j$, either $a_{jl}=0$\ or $\sum_{k\neq l}a_{jk}x_{k}=0$. \ Intuitively,
once we sneak the $2$-norm in \textquotedblleft through the back
door\textquotedblright\ in defining the norm of a complex number, consistency
forces us to use it everywhere. \ We omit a proof of this fact, since it
follows easily from a case analysis similar to that for real amplitudes.

Stepping back, what can we say about why the $2$-norm is special? \ The
standard answer---that the $2$-norm is special because it's preserved under
rotations---merely pushes the question from quantum mechanics back to the
Pythagorean Theorem. \ The latter might be thought a good enough place to
stop. \ However, although the Pythagorean Theorem dates back some 3800 years,
I confess to having never \textit{understood} it at a gut level. \ (Have you?)
\ So if pressed, I'd instead answer the question as follows: values of $p$
other than positive even integers are almost nonstarters, since we want
$\left\vert x\right\vert ^{p}$\ to be defined and smooth at $x=0$. \ But when
$p=4,6,8,\ldots$, Equation \ref{porthog}\ involves terms of\ the form $\left(
a_{jk}x_{k}\right)  ^{q}\left(  a_{jl}x_{l}\right)  ^{p-q}$\ where $q$ and
$p-q$ are both positive even integers, and that immediately forces $A$ to be
generalized diagonal. \ So all that's left is $p=2$.

If you still want to define quantum mechanics using a $p$-norm where $p\neq2$,
the only option seems to be \textit{manual normalization}. \ This means that
when a state $\left\vert \psi\right\rangle =\sum_{x}\alpha_{x}\left\vert
x\right\rangle $\ is measured in the standard basis, the probability of
outcome $\left\vert x\right\rangle $\ is $\left\vert \alpha_{x}\right\vert
^{p}/\sum_{y}\left\vert \alpha_{y}\right\vert ^{p}$. \ Since keeping
$\left\vert \psi\right\rangle $\ normalized is no longer imperative, three
options present themselves for how $\left\vert \psi\right\rangle $\ evolves:

\begin{enumerate}
\item[(i)] As usual, $\left\vert \psi\right\rangle $\ can be mapped to
$U\left\vert \psi\right\rangle $\ where $U$ is any unitary matrix.

\item[(ii)] $\left\vert \psi\right\rangle $\ can be mapped to $A\left\vert
\psi\right\rangle $\ where $A$ is any invertible matrix.

\item[(iii)] $\left\vert \psi\right\rangle $\ can be mapped to $A\left\vert
\psi\right\rangle $, but then \textit{local normalization} is performed on the
subsystem acted on by $A$.
\end{enumerate}

To illustrate option (iii), suppose the nonunitary gate%
\[
\left(
\begin{array}
[c]{cc}%
q & r\\
s & t
\end{array}
\right)
\]
is applied to the second qubit of the normalized state $\alpha\left|
00\right\rangle +\beta\left|  01\right\rangle +\gamma\left|  10\right\rangle
+\delta\left|  11\right\rangle $. \ Then the unnormalized result is%
\[
\left(  q\alpha+r\beta\right)  \left|  00\right\rangle +\left(  s\alpha
+t\beta\right)  \left|  01\right\rangle +\left(  q\gamma+r\delta\right)
\left|  10\right\rangle +\left(  s\gamma+t\delta\right)  \left|
11\right\rangle ,
\]
so the locally normalized result is%
\[
\frac{\sqrt{\alpha^{2}+\beta^{2}}\left[  \left(  q\alpha+r\beta\right)
\left|  00\right\rangle +\left(  s\alpha+t\beta\right)  \left|
01\right\rangle \right]  }{\sqrt{\left(  q\alpha+r\beta\right)  ^{2}+\left(
s\alpha+t\beta\right)  ^{2}}}+\frac{\sqrt{\gamma^{2}+\delta^{2}}\left[
\left(  q\gamma+r\delta\right)  \left|  10\right\rangle +\left(
s\gamma+t\delta\right)  \left|  11\right\rangle \right]  }{\sqrt{\left(
q\gamma+r\delta\right)  ^{2}+\left(  s\gamma+t\delta\right)  ^{2}}},
\]
in contrast to the globally normalized result of%
\[
\frac{\left(  q\alpha+r\beta\right)  \left|  00\right\rangle +\left(
s\alpha+t\beta\right)  \left|  01\right\rangle +\left(  q\gamma+r\delta
\right)  \left|  10\right\rangle +\left(  s\gamma+t\delta\right)  \left|
11\right\rangle }{\sqrt{\left(  q\alpha+r\beta\right)  ^{2}+\left(
s\alpha+t\beta\right)  ^{2}+\left(  q\gamma+r\delta\right)  ^{2}+\left(
s\gamma+t\delta\right)  ^{2}}}.
\]

So, what's wrong with these prescriptions? \ Nothing, as long as you can
stomach the following:

\textbf{(1) Distinguishability of non-orthogonal states.} \ Here's how to
distinguish $d=\Omega\left(  \sqrt{p}\right)  $\ states of a single qubit with
constant probability of error, under option (i) (and therefore under (ii) and
(iii) as well). \ Let the $j^{th}$\ state be $\left\vert \psi_{j}\right\rangle
=\cos\left(  \pi j/d\right)  \left\vert 0\right\rangle +\sin\left(  \pi
j/d\right)  \left\vert 1\right\rangle $ where $j\in\left\{  0,\ldots
,d-1\right\}  $. \ Apply a $d\times d$\ unitary matrix to $\left\vert \psi
_{j}\right\rangle $\ whose first two columns are%
\[
\left(
\begin{array}
[c]{c}%
\cos\left(  \pi0/d\right)  /\sqrt{d}\\
\vdots\\
\cos\left(  \pi\left(  d-1\right)  /d\right)  /\sqrt{d}%
\end{array}
\right)  ,\left(
\begin{array}
[c]{c}%
\sin\left(  \pi0/d\right)  /\sqrt{d}\\
\vdots\\
\sin\left(  \pi\left(  d-1\right)  /d\right)  /\sqrt{d}%
\end{array}
\right)  .
\]
Then measure in the standard basis. \ Suppose without loss of generality that
$j=0$ and that $d$ is odd; then the probability of any outcome other than $0$
being measured is $q/\left(  q+1\right)  $\ where%
\begin{align*}
q  &  =2\sum_{k=1}^{\left(  d-1\right)  /2}\left\vert \cos\left(  \frac{\pi
k}{d}\right)  \right\vert ^{p}\\
&  \leq2\sum_{k=1}^{\left(  d-1\right)  /2}\left(  1-\frac{\left(  \pi
k/d\right)  ^{2}}{2}+\frac{\left(  \pi k/d\right)  ^{4}}{24}\right)  ^{p}\\
&  \leq2\sum_{k=1}^{\left(  d-1\right)  /2}\left(  1-\frac{\pi^{2}k^{2}%
}{4d^{2}}\right)  ^{2p}\\
&  \leq2\sum_{k=1}^{\left(  d-1\right)  /2}\exp\left(  -\frac{\pi^{2}k^{2}%
p}{2d^{2}}\right)
\end{align*}
which is bounded away from $1$ so long as $p\geq cd^{2}$\ for some constant
$c$. \ It would be interesting to obtain bounds on how many states can be
reliably distinguished in higher-dimensional Hilbert spaces.

\textbf{(2) Superluminal signalling.} \ Under option (ii), given an EPR pair,
Alice can communicate a bit to Bob by mapping $\left\vert 00\right\rangle
+\left\vert 11\right\rangle $\ to either $\left\vert 00\right\rangle
+\varepsilon\left\vert 11\right\rangle $\ or $\varepsilon\left\vert
00\right\rangle +\left\vert 11\right\rangle $. \ Indeed, using the ideas from
part (1), she can communicate $\Omega\left(  \sqrt{p}\right)  $\ bits to Bob
using a single EPR pair! \ I conjecture this is tight. \ Under options (i) and
(iii), Alice can communicate a bit to Bob given enough EPR pairs, by taking
advantage of Bob's ability to distinguish nonorthogonal states. \ Note that
under options (ii) and (iii), superluminal signalling is possible even when
$p=2$.

\textbf{(3) Efficient solubility of }$\mathsf{NP}$\textbf{-complete and even
harder problems.} \ Suppose you're given a Boolean function $f:\left\{
0,1\right\}  ^{n}\rightarrow\left\{  0,1\right\}  $. \ Under option (ii),
first prepare $\sum_{x}\left\vert x\right\rangle \left\vert f\left(  x\right)
\right\rangle $, then apply the nonunitary gate%
\begin{equation}
G=\left(
\begin{array}
[c]{cc}%
2^{-2n} & 0\\
0 & 1
\end{array}
\right)  \label{gateg}%
\end{equation}
to the $f$ register and measure to learn whether there exists an $x$ such that
$f\left(  x\right)  =1$. \ Indeed, Section \ref{POSTBQP} shows that under
options (i), (ii), and (iii), you could solve even $\mathsf{PP}$-complete
problems in polynomial time, which are believed to be harder than
$\mathsf{NP}$-complete problems.

\textbf{(4) Singularity.} \ Under options (ii) and (iii), the matrix $A$ could
be arbitrarily close to a non-invertible matrix, which can map nonzero states
to the zero state.

\section{Quantum Computing With Postselection\label{POSTBQP}}

This section can be skipped by physicists with no interest in computational
complexity.\footnote{The rest of paper can be skipped by computational
complexity theorists with no interest in physics.} \ Its goal is to show that,
if you change quantum mechanics in any of three ways, then the class of
problems efficiently solvable on a quantum computer expands drastically, from
$\mathsf{BQP}$\ (Bounded-Error Quantum Polynomial-Time) to $\mathsf{PP}%
$\ (Probabilistic Polynomial-Time). \ Here $\mathsf{PP}$\ is a well-studied
classical complexity class, consisting of all decision problems for which
there exists a probabilistic polynomial-time Turing machine that accepts with
probability at least $1/2$\ if the answer is \textquotedblleft
yes,\textquotedblright\ and with probability less than $1/2$\ if the answer is
\textquotedblleft no.\textquotedblright\ \ The three changes that would give
quantum computers the power of $\mathsf{PP}$\ are: replacing the $2$-norm by
the $p$-norm for any $p\neq2$,\footnote{If $p$ is not a positive even integer,
then the power increases \textit{at least} to $\mathsf{PP}$ and possibly
further.} allowing arbitrary invertible matrices instead of just unitary
matrices, or allowing postselection on measurement outcomes. \ Any
\textit{combination} of these changes would also yield $\mathsf{PP}$. \ Note,
however, that I always assume global normalization (corresponding to options
(i) and (ii) in Section \ref{PNORM}).

It will be convenient to define a new complexity class:

\begin{definition}
\label{postbqpdef}$\mathsf{PostBQP}$ (or $\mathsf{BQP}$\ with postselection)
is the class of languages $L$ for which there exists a uniform family of
polynomial-size quantum circuits such that for all inputs $x$,

\begin{enumerate}
\item[(i)] At the end of the computation, the first qubit has a nonzero
probability of being measured to be $\left\vert 1\right\rangle $.

\item[(ii)] If $x\in L$, then conditioned on the first qubit being $\left\vert
1\right\rangle $, the second qubit is $\left\vert 1\right\rangle $\ with
probability at least $2/3$.

\item[(iii)] If $x\notin L$, then conditioned on the first qubit being
$\left\vert 1\right\rangle $, the second qubit is $\left\vert 1\right\rangle
$\ with probability at most $1/3$.
\end{enumerate}
\end{definition}

Intuitively, postselection\ means that at some point in the computation, you
can measure a qubit that has a nonzero probability of being $\left\vert
1\right\rangle $, and \textit{assume} that the outcome will be $\left\vert
1\right\rangle $\ (or equivalently, discard all runs where the outcome is
$\left\vert 0\right\rangle $). \ Just as Bernstein and Vazirani \cite{bv}%
\ showed that intermediate measurements don't increase the power of ordinary
quantum computers, so it's easy to show that intermediate postselection steps
don't increase the power of $\mathsf{PostBQP}$ (since these steps can all be
deferred to the end). \ On the other hand, if operations can be performed
conditioned on measurement outcomes, then mixing postselection and measurement
steps \textit{could} increase the power of $\mathsf{PostBQP}$.

In the remainder of the section, I'll first show that $\mathsf{PostBQP}%
=\mathsf{PP}$\ (Theorem \ref{postbqppp}), and then use that result to prove
that the other changes also give quantum computers the power of $\mathsf{PP}$.

\begin{theorem}
\label{postbqppp}$\mathsf{PostBQP}=\mathsf{PP}$.
\end{theorem}

\begin{proof}
The inclusion $\mathsf{PostBQP}\subseteq\mathsf{PP}$ follows easily from the
techniques used by Adleman, DeMarrais, and Huang \cite{adh}\ to show that
$\mathsf{BQP}\subseteq\mathsf{PP}$.

For the other direction, let $f:\left\{  0,1\right\}  ^{n}\rightarrow\left\{
0,1\right\}  $\ be a Boolean function and let $s=\left\vert \left\{
x:f\left(  x\right)  =1\right\}  \right\vert $. \ Then we need to decide in
$\mathsf{PostBQP}$\ whether $s<2^{n-1}$ or $s>2^{n-1}$. \ (As a technicality,
we can guarantee using padding that $s>0$ and $s\neq2^{n-1}$.) \ The algorithm
is as follows: first prepare $2^{-n/2}\sum_{x\in\left\{  0,1\right\}  ^{n}%
}\left\vert x\right\rangle \left\vert f\left(  x\right)  \right\rangle $.
\ Then following Abrams and Lloyd \cite{al}, apply Hadamard gates to all $n$
qubits in the first register and postselect\footnote{Actually postselection is
overkill here, since the first register has at least $1/4$ probability of
being $\left\vert 0\right\rangle ^{\otimes n}$.} on that register being
$\left\vert 0\right\rangle ^{\otimes n}$,\ to obtain $\left\vert
0\right\rangle ^{\otimes n}\left\vert \psi_{s}\right\rangle $\ where%
\[
\left\vert \psi_{s}\right\rangle =\frac{\left(  2^{n}-s\right)  \left\vert
0\right\rangle +s\left\vert 1\right\rangle }{\sqrt{\left(  2^{n}-s\right)
^{2}+s^{2}}}.
\]
Next, for some positive real $\alpha,\beta$\ to be specified later, prepare
$\alpha\left\vert 0\right\rangle \left\vert \psi_{s}\right\rangle
+\beta\left\vert 1\right\rangle \left\vert \phi_{s}\right\rangle $ where
\[
\left\vert \phi_{s}\right\rangle =\frac{2^{n}\left\vert 0\right\rangle
+\left(  2^{n}-2s\right)  \left\vert 1\right\rangle }{\sqrt{2\left(
2^{n}-s\right)  ^{2}+2s^{2}}}%
\]
is the result of applying a Hadamard to $\left\vert \psi_{s}\right\rangle $.
\ Postselecting on the second qubit being $\left\vert 1\right\rangle $\ then
yields the state%
\[
\left\vert \varphi_{s,\beta/\alpha}\right\rangle =\frac{\alpha s\left\vert
0\right\rangle +\sqrt{1/2}\beta\left(  2^{n}-2s\right)  \left\vert
1\right\rangle }{\sqrt{\alpha^{2}s^{2}+\beta^{2}\left(  2^{n}-2s\right)
^{2}/2}}%
\]
in the first qubit. \ A simple calculation now reveals that if $s<2^{n-1}$,
then there exists an integer $i$\ in the range $\left[  -n,n\right]  $\ such
that%
\[
\left\vert \left\langle +|\varphi_{s,2^{i}}\right\rangle \right\vert
\geq\frac{1+\sqrt{2}}{\sqrt{6}}\approx0.986
\]
where $\left\vert +\right\rangle =\left(  \left\vert 0\right\rangle
+\left\vert 1\right\rangle \right)  /\sqrt{2}$.\ \ If $s>2^{n-1}$,\ on the
other hand, then for all such $i$\ we have $\left\vert \left\langle
+|\varphi_{s,2^{i}}\right\rangle \right\vert \leq1/\sqrt{2}$. \ So by running
the whole algorithm $n\left(  2n+1\right)  $\ times in parallel, with
$n$\ invocations for each integer $i\in\left[  -n,n\right]  $, we can learn
whether $s<2^{n-1}$ or $s>2^{n-1}$\ with exponentially small probability of error.
\end{proof}

Let $\mathsf{BQP}_{\text{\textsf{nu-global}}}$\ be the class of problems
solvable by a uniform family of polynomial-size, bounded-error quantum
circuits, if the circuits can consist of arbitrary invertible gates, not just
unitary gates. \ Option (ii) from Section \ref{PNORM}\ is used for
normalization; that is, before a measurement, the amplitude $\alpha_{x}$\ of
each basis state $\left\vert x\right\rangle $\ is divided by $\sqrt{\sum
_{y}\left\vert \alpha_{y}\right\vert ^{2}}$.

\begin{proposition}
\label{nuglobal}$\mathsf{BQP}_{\text{\textsf{nu-global}}}=\mathsf{PP}$.
\end{proposition}

\begin{proof}
The inclusion $\mathsf{BQP}_{\text{\textsf{nu-global}}}\subseteq\mathsf{PP}%
$\ follows easily from Ref. \cite{adh}. \ For the other direction, by Theorem
\ref{postbqppp}\ it suffices to observe that $\mathsf{PostBQP}\subseteq
\mathsf{BQP}_{\text{\textsf{nu-global}}}$. \ To postselect on a qubit being
$\left\vert 1\right\rangle $, simply apply the nonunitary gate $G$ from
Equation \ref{gateg}.
\end{proof}

Define $\mathsf{BQP}_{\text{\textsf{nu-local}}}$\ similarly to $\mathsf{BQP}%
_{\text{\textsf{nu-global}}}$, except that after every gate $G$, option (iii)
(local normalization) is applied to the qubits acted on by $G$. \ Assume that
arbitrary $1$- and $2$-qubit gates are available to polynomially many bits of precision.

\begin{proposition}
\label{nulocal}$\mathsf{PP}\subseteq\mathsf{BQP}_{\text{\textsf{nu-local}}%
}\subseteq\mathsf{PSPACE}$.
\end{proposition}

\begin{proof}
For $\mathsf{PP}\subseteq\mathsf{BQP}_{\text{\textsf{nu-local}}}$, observe
that in the proof of Theorem \ref{postbqppp}, the only essential postselection
steps are applied to $2$-qubit pure states unentangled with anything else.
\ For $2$-qubit gates acting on these states, local normalization is the same
as global normalization.

For $\mathsf{BQP}_{\text{\textsf{nu-local}}}\subseteq\mathsf{PSPACE}$, let
$\alpha_{x}^{\left(  t\right)  }$\ be the amplitude of basis state $\left\vert
x\right\rangle $\ at time $t$. \ Then for all $x,t$\ we can write $\alpha
_{x}^{\left(  t\right)  }$ as a function of $\alpha_{y_{1}}^{\left(
t-1\right)  },\ldots,\alpha_{y_{k}}^{\left(  t-1\right)  }$\ for some constant
$k$ and basis states $\left\vert y_{1}\right\rangle ,\ldots,\left\vert
y_{k}\right\rangle $. \ This immediately implies a depth-first recursive
algorithm (using a polynomial amount of memory) for approximating any
amplitude $\alpha_{x}^{\left(  t\right)  }$ to polynomially many bits of precision.
\end{proof}

Finally, for any nonnegative real number $p$, define $\mathsf{BQP}_{p}%
$\ similarly to $\mathsf{BQP}$, except that the probability of measuring a
basis state $\left\vert x\right\rangle $\ equals $\left\vert \alpha
_{x}\right\vert ^{p}/\sum_{y}\left\vert \alpha_{y}\right\vert ^{p}$. \ (Thus
$\mathsf{BQP}_{2}=\mathsf{BQP}$.) \ All gates are unitary.

\begin{proposition}
\label{bqpp}$\mathsf{PP}\subseteq\mathsf{BQP}_{p}\subseteq\mathsf{P}%
^{\mathsf{\#P}}$ for all constants $p\neq2$, and $\mathsf{BQP}_{p}%
=\mathsf{PP}$ provided $p$ is an even integer greater than $2$.
\end{proposition}

\begin{proof}
The inclusion $\mathsf{BQP}_{p}\subseteq\mathsf{P}^{\mathsf{\#P}}$\ is
obvious. \ To simulate $\mathsf{BQP}_{p}$ in $\mathsf{PP}$ when $p$ is a
positive even integer, use the techniques of Ref. \cite{adh} (which handle the
$p=2$ case), but evaluate polynomials of degree $p$ instead of quadratic
polynomials. \ To simulate $\mathsf{PP}$ in $\mathsf{BQP}_{p}$\ when $p\neq2$,
run the algorithm of Theorem \ref{postbqppp},\ having initialized $O\left(
n^{3}/\left\vert p-2\right\vert \right)  $ ancilla qubits to $\left\vert
0\right\rangle $. \ To postselect on the $b^{th}$ qubit being $\left\vert
1\right\rangle $: if $p<2$, then\ apply Hadamards to $10pn/\left(  2-p\right)
$\ ancilla qubits conditioned on the $b^{th}$\ qubit being $\left\vert
1\right\rangle $. \ If $p>2$, then apply Hadamards to $10pn/\left(
p-2\right)  $\ ancilla qubits conditioned on the $b^{th}$\ qubit being
$\left\vert 0\right\rangle $. \ 
\end{proof}

\section{Real Amplitudes\label{REAL}}

\begin{quote}
\textquotedblleft C'mon, they're algebraically closed!\textquotedblright\ ---A
math graduate student, when asked why God would resort to complex numbers in
creating quantum mechanics
\end{quote}

To a beginner, perhaps the most unexpected fact about quantum mechanics is
that amplitudes are complex. \ As the term `imaginary' suggests, we tend to
think of complex numbers as (useful) human inventions;\ it's unsettling if the
source code of the Universe is best written in a language like Fortran with a
complex-number data type. \ Also, in contrast to what we saw in Section
\ref{PNORM}, restricting amplitudes to be real doesn't lead to a theory
obviously very different from quantum mechanics. \ All the greatest hits are
still there: interference, entanglement, Bell inequality violations,
noncommuting observables, non-unique decompositions of mixed states, universal
quantum computing, the Zeno effect, the Gleason and Kochen-Specker theorems.

Nevertheless, Section \ref{INTRO}\ recalled a subtle difference between
complex and real (or for that matter complex and quaternionic) amplitudes,
based on counting the number of parameters of a mixed state. \ This section
gives a completely different argument for why amplitudes aren't real. \ The
advantage of this argument is that it's elementary and intuitive; the
disadvantage is that it says nothing about why amplitudes are complex rather
than quaternionic.

Let $\mathcal{S}$ be a set of states, and let $\mathcal{U}$ be a set of
transformations from $\mathcal{S}$ to itself. \ Say $\mathcal{U}$ has the
\textit{square root property} if for all $U\in\mathcal{U}$, there exists
another transformation $V\in\mathcal{U}$\ such that $V\left(  V\left(
S\right)  \right)  =U\left(  S\right)  $\ for all $S\in\mathcal{S}$. \ If time
is continuous, then the importance of the square root property is obvious:
without it there are transformations that can't be interpreted as the result
of applying a fixed Hamiltonian for some interval of time. \ Even if time is
discrete, the square root property is desirable, because it allows any
$U$\ that acts over $k$ time steps to be approximated by $V^{k}$\ for some $V$
that acts over a single time step.\footnote{To write $U$ exactly as $V^{k}%
$\ we'd need \textquotedblleft a $k^{th}$\ root property,\textquotedblright
\ which also holds for quantum mechanics but fails for real quantum
mechanics.} \ Clearly quantum mechanics has the square root property: given a
unitary $U$, let $\left\vert \psi_{1}\right\rangle ,\ldots,\left\vert \psi
_{n}\right\rangle $\ be the eigenvectors of $U$ and let $\lambda_{1}%
,\ldots,\lambda_{n}$\ be the corresponding eigenvalues; then there exists a
unitary $V$ with eigenvectors $\left\vert \psi_{1}\right\rangle ,\ldots
,\left\vert \psi_{n}\right\rangle $\ and eigenvalues $\mu_{1},\ldots,\mu_{n}%
$\ such that $\mu_{j}^{2}=\lambda_{j}$, which therefore satisfies $V^{2}=U$.
\ Since every quaternion has a square root,\footnote{Indeed some, such as
$-1$, have infinitely many square roots.} the same argument shows that
quaternionic quantum mechanics has the square root property.

However, real quantum mechanics doesn't have the square root property. \ This
is immediate since orthogonal matrices such as%
\[
\left(
\begin{array}
[c]{cc}%
1 & 0\\
0 & -1
\end{array}
\right)  ,\left(
\begin{array}
[c]{cc}%
0 & 1\\
1 & 0
\end{array}
\right)
\]
with determinant $-1$\ can't be written as squares of matrices with real
determinants. \ If we want to restore the square root property, then we have
two choices. \ The first choice is to restrict to the group
$\operatorname*{SO}\left(  n\right)  $---that is, to real orthogonal matrices
with determinant $1$. \ It's not hard to see that for every $U\in
\operatorname*{SO}\left(  n\right)  $, there exists a $V\in\operatorname*{SO}%
\left(  n\right)  $\ such that $V^{2}=U$. \ On the other hand, natural
$1$-qubit operations such as the above two can only be implemented by using
ancillia qubits. \ The second choice is to allow the \textquotedblleft square
root\textquotedblright\ of $U$ to have larger dimension than $U$. \ For
example, \
\[
\left(
\begin{array}
[c]{ccc}%
1 & 0 & 0\\
0 & 0 & 1\\
0 & -1 & 0
\end{array}
\right)  ^{2}=\left(
\begin{array}
[c]{ccc}%
1 & 0 & 0\\
0 & -1 & 0\\
0 & 0 & -1
\end{array}
\right)
\]
contains the\ $1$-qubit phase flip as a $2\times2$\ submatrix. \ This is an
instance of a well-known geometrical fact, that a mirror reversal in $n$
dimensions can be accomplished by a rotation in $n+1$\ dimensions. \ Indeed,
\textit{any} $n\times n$\ orthogonal matrix $U$ has a real square root of
dimension $\left(  n+1\right)  \times\left(  n+1\right)  $, since there exists
an element of $\operatorname*{SO}\left(  n+1\right)  $ that contains $U$ as a
submatrix. \ With either choice, the price we pay is that our $n$-dimensional
theory can be fully described only in $n+1$\ dimensions. \ But the $\left(
n+1\right)  $-dimensional theory requires $n+2$\ dimensions to describe, and
so on ad infinitum---unless we declare that the $\left(  n+1\right)  ^{st}%
$\ dimension\ is physically different from dimensions $1$\ to $n$.

\section*{Acknowledgments}

I thank Andrei Khrennikov for including this paper in the V\"{a}xj\"{o}
proceedings despite its having no relation to anything I talked about at the
conference; and Chris Fuchs, without whom this \textquotedblleft saucy
paper\textquotedblright\ (his words) wouldn't have been written.

\end{document}